\documentclass[conference]{IEEEtran}
\IEEEoverridecommandlockouts
\usepackage{cite}
\usepackage{amsmath,amssymb,amsfonts}
\usepackage{algorithmic}
\usepackage{graphicx}
\usepackage{textcomp}
\usepackage{xcolor}
\usepackage{graphicx}
\usepackage{wrapfig}
\usepackage{float}
\graphicspath{ {./images/} }
\def\BibTeX{{\rm B\kern-.05em{\sc i\kern-.025em b}\kern-.08em
    T\kern-.1667em\lower.7ex\hbox{E}\kern-.125emX}}
\begin{document}

\title{Music Generation Using Three-layered LSTM}

\makeatletter
\newcommand{\linebreakand}{%
  \end{@IEEEauthorhalign}
  \hfill\mbox{}\par
  \mbox{}\hfill\begin{@IEEEauthorhalign}
}
\makeatother

\author{
\IEEEauthorblockN{1\textsuperscript{st} Vaishali Ingale}
\IEEEauthorblockA{
\textit{Army Institute of Technology}\\
Pune, India \\
vingale@aitpune.edu.in}
\and
\IEEEauthorblockN{2\textsuperscript{nd} Anush Mohan}
\IEEEauthorblockA{
\textit{Army Institute of Technology}\\
Pune, India \\
anushmohan\_17380@aitpune.edu.in}
\and
\IEEEauthorblockN{3\textsuperscript{rd} Divit Adlakha}
\IEEEauthorblockA{
\textit{Army Institute of Technology}\\
Pune, India \\
divitadlakha\_17493@aitpune.edu.in}
\linebreakand
\IEEEauthorblockN{4\textsuperscript{th} Krishan Kumar}
\IEEEauthorblockA{
\textit{Army Institute of Technology}\\
Pune, India \\
krishnakumar\_17564@aitpune.edu.in}
\and
\IEEEauthorblockN{5\textsuperscript{th} Mohit Gupta}
\IEEEauthorblockA{
\textit{Army Institute of Technology}\\
Pune, India \\
mohitgupta\_17429@aitpune.edu.in}
}

\maketitle

\begin{abstract}
This paper explores the idea of utilising Long Short-Term Memory neural networks (LSTMNN) for the generation of musical sequences in ABC notation. The proposed approach takes ABC notations from the Nottingham dataset and encodes it to be fed as input for the neural networks. The primary objective is to input the neural networks with an arbitrary note, let the network process and augment a sequence based on the note until a good piece of music is produced. Multiple calibrations have been done to amend the parameters of the network for optimal generation. The output is assessed on the basis of rhythm, harmony, and grammar accuracy.

\end{abstract}


\section{Introduction}
It is a myth that you need to be a musician to generate music. Even an music enthusiast can produce a musical piece through the means of technology. 
Until recently, all music generation was done manually
by means of analogue signals. However, now, music
production is done through technology, assisted by humans.
The task that has been accomplished in the paper is the construction of neural network architectures that can
efficiently portray the complex details of harmony and melody
without the need for human intervention. A brief summary of
the precise details of music and its mechanisms has been provided
in the paper with appropriate citations where required. The
primary objective of the papers was to devise an algorithm that
can be used to create musical notes utilizing Long Short-Term
Memory (LSTM) [9] networks in Recurrent Neural Networks
(RNNs) [2]. The output data obtained is in ABC notation.

To train the model we have chosen to work on ABC notations. ABC notation is one of the ways to represent music. It consists of two parts. First part represents the meta data which comprises various characteristics of tune such as index, time signature, default note length and type of tune. And second part represents the actual tune which is nothing but a sequence of characters. The devised algorithm learns the
sequences of monophonic musical notes over three single layered LSTM network.

The following sections cover other work that has been conducted in the field, the details of the approach taken and the subsequent results thus obtained.

\section{Literature Review}

Music generation has been at the epicenter of attention of members of the research community and has thus been studied upon a lot. Many who did try to generate music have done so using different approaches. Hence there exists numerous ways in which music can be generated and an amalgamation of such approaches can be used to create and design a new yet competent model. These approaches have been divided into two main categories – Traditional[4] and Autonomous [3]. Traditional approach uses algorithms working on already defined functions to make music, whereas Autonomous Model learn from the prior iterations of the notations and then generates new ones. 
Algebra founded upon the usage of the tree structure to enforce grammar constitutes one of the earlier attempts [4]. Markov chains[5] can be used to design such a model [2]. 
Many models and approaches have been documented in the field of artificial intelligence soon after the field experienced a massive boom. Such models include probabilistic models which use variants of RNN, namely Char RNNs, Anticipation RNNs [7] [2]. One method that is actively being used to generate musical notes is the Generative adversarial networks (GANs) which contains two neural networks – discriminator network and the generator network [8]. The generator network and the discriminator network work in tandem to evaluate authenticity of the generated data against the original dataset. 
Research studies reveal that LSTM outclasses the GAN in terms of fixating on certain sequences, that is, LSTM is superior when it comes to uncovering specific patterns and then recycling them throughout the course of the output sequence. The models based upon LSTM were able to get out of certain note loops and shift into other notes[9]. When it came to GAN-based models, they were only able to pickup on basic concepts, albeit better, and they exhibited shorter training times [8].

\section{The Methodology}

\subsection{Objectives and Technical Challenges}\label{AA}
The initial challenge faced when dealing with music is that of its representation. Signals, MIDI, notations, etc. are all possible representations. Due to the higher efficacy of notations for the task at hand, the model used in this paper is fed input and generates output in ABC notations. The ABC notation uses 7 letters (A to G) with other symbols representing features such as – flat, sharp, note length, key, etc. to represent the given notes. 

However, the subsequent issue is that if the output is in the form of notations, it is not readily eligible by the majority of humans. To make the output understandable by an amateur, the notations can be easily converted into the preferred audio format. In our testings, we converted the notation-based sequence into MIDI and MP3 formats.  

\subsection{Design}
RNNs fall victim to the vanishing/exploding gradient problem due to their use of back-propagation[2], to remedy this we have employed the use of LSTMs. At each timestep of the RNN, the individual LSTM[9] cell is fed a value, the cell then calculates the hidden vector and outputs is to the next timesteps. 
The current input, and the previous hidden and memory states are given as input to the cell. Similarly, the current hidden state and the current memory state are the outputs. Taking the current timestep as t, the current hidden vector ht is found using the current input at and the previous timestep’s hidden vector ht – 1. This is how RNNs process data sequentially.

\begin{figure}
    \centering
    \includegraphics[width=0.48\textwidth]{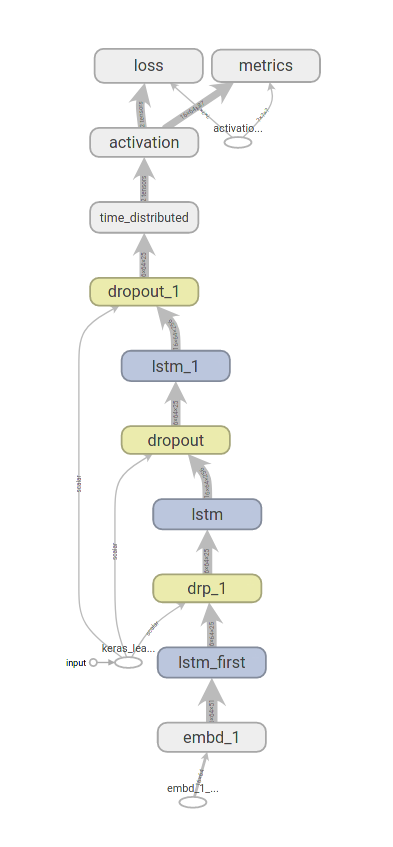}
    \caption{The figure depicts the flow of data through the layers present in the model.}
    \label{fig_1}
\end{figure}

\subsection{Data Processing}

The model is designed to interpret the musical notes in the form of 87 unique characters present in a dictionary, using integer encoding. The dataset is hot-encoded as well, to convert the labels into binary vectors. This is then fed into the LSTM units in batches. 
The specification of the batches used here are as follows: Batch Size = 16, Sequence Length = 64. 

\subsection{Architecture}

The model is built around 3 LSTM [2] layers, acting as the core. Due to the multi-class classification nature of the problem statement, the SoftMax activation function is deployed as well. Dropout layer is also present to aid in the avoidance of overfitting. The Adam optimizer is used since the model deals with RNNs [2]. Also, to process the outputs at each timestep, Time Distributed Dense Layer is utilized here.

\begin{figure}[H]
    \centering
    \includegraphics[width=0.48\textwidth]{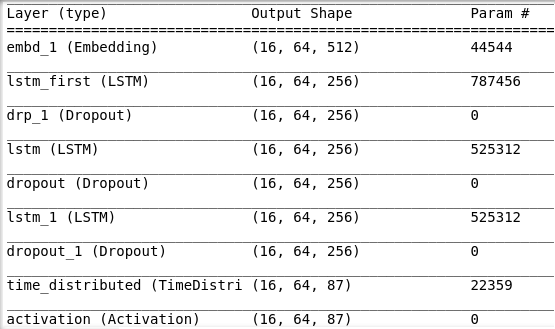}
    \caption{The figure shows the architecture, that is the different layers, their input and output sizes etc.}
    \label{fig_2}
\end{figure}

\section{Results \& Discussion}
The model is implemented using 90 epochs which yields in less training loss per epoch with increasing number of epochs which in turn result in training accuracy of 94\%. Values of some epoch’s loss and accuracy are shown in the table [Fig3].

Tensorboard outputs (visualizations) represents these values in the form of graphs. First graph [Fig4] describes the training accuracy values of the model with corresponding epochs. The initial training accuracy level is observed to be 59\%, which then linearly increases 5th epoch onwards, that is from 75\% to 94\% linear increase in span of 85 epochs.

The second graph [Fig5] have variables epoch and its corresponding training loss. Here, the gradual reduction in training loss is clearly visible with higher number of epochs. With 1.4359 training loss in the 1st epoch to the final 0.1737 loss at the last epoch, the reduction can clearly be seen in Fig. 3.

\begin{figure}[H]
    \centering
    \includegraphics{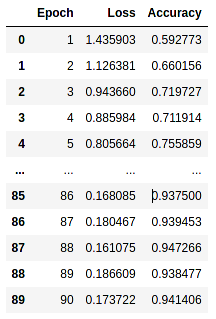}
    \caption{The table is the output of running the model for 90 epochs, yielding upto 94\% accuracy.}
    \label{fig_3}
\end{figure}

\begin{figure}[H]
    \centering
    \includegraphics[width=0.48\textwidth]{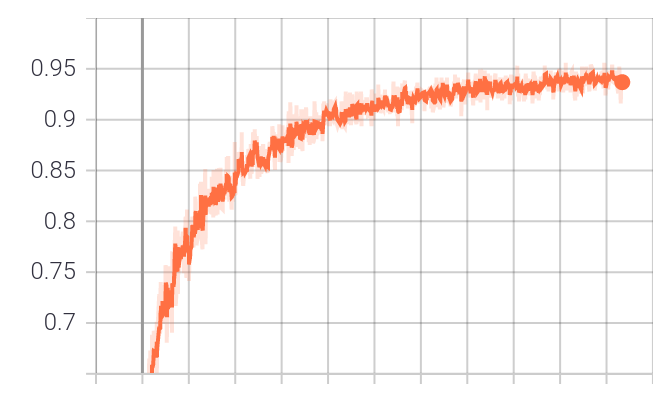}
    \caption{The graph represents the Tensorboard output of the training accuracy over time. X-Axis: Time Steps. Y-Axis: Accuracy}
    \label{fig_4}
\end{figure}

\begin{figure}[H]
    \centering
    \includegraphics[width=0.48\textwidth]{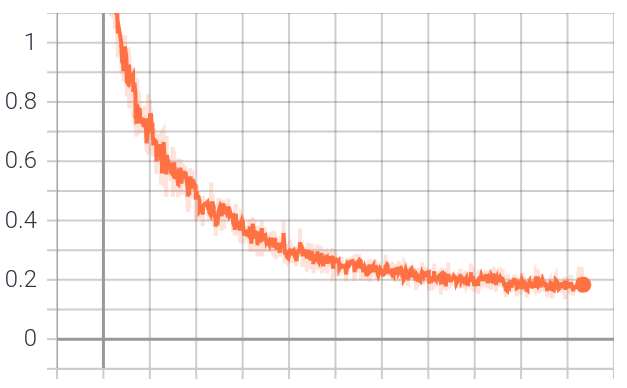}
    \caption{The graph represents the Tensorboard output of the training loss over time. X-Axis: Time Steps. Y-Axis: Loss}
    \label{fig_5}
\end{figure}

\section{Conclusions}
LSTM[9] RNNs propose a promising approach for automated sequence generation. These networks excel at predicting the next member of the sequence by making decisions based on context. The saving of weights during back propagation during each epoch has led to greater accuracy and lower losses. The performance of this model can be more impressive if the dataset is altered to include more tunes, in variety, and that of multiple instruments. The training, if done more rigorously, can yield better results as well.

The monophonic music generated here is pleasant to the ear and has high accuracy. Training the model with polyphonic data can help make the output sequence more appreciable to the normal person.\\

The model is trained on monophony, that is, the dataset is comprised of a single instrument. Further progressions can be made by delving into multiple instruments, that is, polyphony.

The dataset can also be further expanded to comprise of more tunes, in variety, so that the model can have a more robust training.

\end{document}